# *Digitory*, A Smart Way of Learning Islamic History in Digital Era

**Dimas Agil Marenda, Ahmad Nasikun, Canggih Puspo Wibowo**

Department of Electrical Engineering and Information Technology, Faculty of Engineering,

Universitas Gadjah Mada (UGM), Yogyakarta, Indonesia

*Email:* dimasagil@te.gadjahmada.edu; ahmad.nasikun@ugm.ac.id;

canggih@te.gadjahmada.edu

**Abstract:** As a country with the greatest Muslim number, Indonesia has a long history on its introduction to Islam. Unfortunately, there has not been effective and interactive ways of learning Islamic history in Indonesia, beside text-books-based stories, which may not be interesting for students. Therefore, Indonesian Muslim students may not value Islamic history, so that they cannot learn from and implement it in their lives.

On other side, Information and Communication Technology (ICT) has grown rapidly in the last few decades, bringing new means for people in conducting their tasks. One of the most fast-growing technologies is web-based application that can transmit data worldwide on ease. Seemingly web has been an interactive way of communicating and disseminating information among web users, including to counter problem of Islamic history learning as given previously.

In this paper, we would like to propose our web-based application for Islamic history learning system, named Digitory—Digital History. There are two main navigations to browse historical stories on the web. One is location-based navigation using on-line map which gives users access to understand events during Islamic introduction to Indonesia based on the location. Alternatively, users can browse on time base using sliding navigation marked timely, so they can comprehend history in time sequence. Implementing semantic web, this web-app can suggest users related articles, either based on time series, location similarities, or time-domain similarities, and also suggest pictures related to the articles. AJAX strengthen the real-time access and interactivity as users navigate around the web.

By the end of this research, we will have the web-app prototype completed with its fundamental location-based navigation, semantic relationship among articles and pictures, and time-based navigation system. As the web-app deployed, we have some volunteers to try our application so that we can evaluate its result, particularly its effectiveness.

## Introduction

History is episodes of humans life in the past. It contains many dynamic aspects that make life in the past becoming inseparable part if current human life and even human life scenario in the past. History also enables human to repeat the glories in the past (Joseph, 2008).

Indonesia is a country with the most number of Muslim population, in which most Indonesians believe in Islam. Islam did not suddenly spread out in Indonesia within a night, but it was preached in decades until it is widely accepted by most Indonesians. Islam has also played important part during Indonesian development, and therefore, learning Islam history in Indonesia is significant part of understanding Islam in Indonesia (Suharto, 2008).

Islam history is very important part of education teaching, particularly for its young generation in facing global challenge in modern Era. Therefore, it is very necessary to provide effective and interactive ways of learning Islamic History, so that youth could learn Islamic values from history and implement it on their daily life.

On the other hand, Information and Communication Technology (ICT) has grown rapidly in the last few decades, bringing new means for people in conducting their tasks. Through VoIP people from different parts of the worlds can talk face-to-face. Products advertisement and distribution are now much easier to conduct with the help of ICTs.

One of the most fast-growing technologies is web, especially its web-based application that can transmit data worldwide on ease. Seemingly web has been an interactive way of communicating and disseminating information among web users, including to counter problem of Islamic history learning as given previously.

**Literature Review**

Digitory is built with open source softwares and standards. It is developed using CodeIgniter PHP Framework with the help of jQuery javascript library. CodeIgniter helps developers create the web because it has many built-in libraries so that developers do not need to code every line from scratch (EllisLab, 2011). Its MVC (Model-View-Controller) design pattern separates data, interface, and business logic within application, enabling developers to focus on one aspect without having to worry to other aspects.

Jquery is javascript library that helps developers to write javascript codes for their web development. Jquery makes developers much easier in using javascript, particularly for user interaction and interface (jQuery, 2011). In Digitory, jQuery is very important in creating time-based navigation and locatin-based navigation, as well as in gallery.

**Navigation**

To have good quality application, it has to provide good user friendly feautures for its users. User friendly generally means the ability of software or application to be easily used and accessed by users so that they are willing to spend time in operating that particular application (Santosa, 2007). Navigation is one key feature in creating good user friendly web application.

Navigation plays very important role within a web site because it allows users to go and jump to various parts of the web easily. It has analogical function with table of contents within text books (Stocks, 2009).

**Semantic web**

The founder of world wide web, Tim Berners-Lee, defines semantic web as "a web of data that can be processed directly and indirectly by machines" (Berners-Lee, 2001). Furthermore, web needs to create smart data so that it can be transferred and understood by other machines. With the emerging of web, Extensible Markup Language (XML), and Semantic web, the role of data within application is much more important than ever (Daconta, 200).

**System Design**

The main features of Digitory is its very user friendly navigation on its homepage, location-based and time-based navigation, that helps users browse its Islamic history contents.

To create good web application, we need to have good system design as well. One important aspect of design is to understand the data flow in the web application, in addition to provide visualisation of final product which plays guideline role (Fathnan, 2010). There are two components in designing, which are content and user interface design.

1. Content Design

Content is the core of Digitory, in which Islamic history in Indonesia is provided to the users. Data in Digitory is timely-bonded, since Islam first introduced in Indonesia until the independence of Indonesia, particularly Islam role during the struggle for independence. Digitory is in Bahasa Indoensia since its target is for education for Indonesian Muslims.

How articles organized is shown in Figure 1 below. Islamic history data are described in articles, which are grouped under "Glossary". Format for all articles is standardized in similar patterns: image thumbnails, main article content, related images, article structures, links to related articles, and sharing to social networks. Glossary in this term means group of historical articles with similar themes and close time periods.

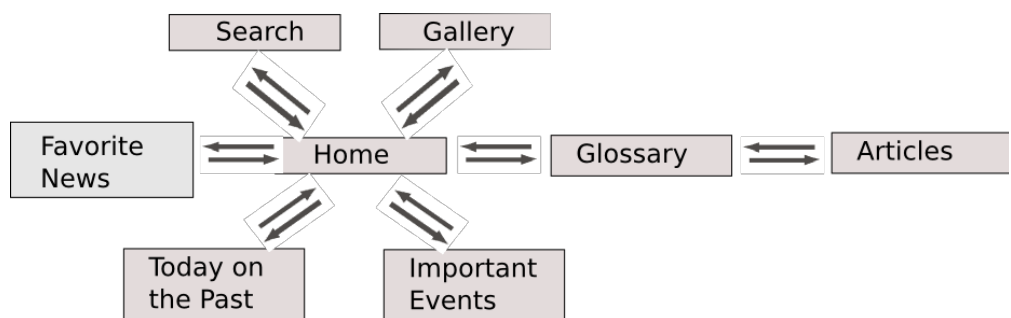

Figure 1 . Content Scheme of Digitory

Digitory is also equipped with several additional features, including "Today in the Past" that will smartly render articles whose events happen in the past but on the same date with the day user visits the web. Gallery helps users to understand events by showing users images of related to it.

2. User Interface

User interface (UI) plays very important role in interaction between user and application programs. Easy and friendly user interface makes it easier for users to browse the contents, which is Islamic history. In designing UI within a web application, navigation plays very important role, to make users have easy access to wherever page they want.

In homepage, Digitory menus provide links to content of articles, including home, glossary, gallery, important events, and searching. Using dynamic navigation, different submenus are rendered for different menu clicked. Time-based navigation is divided into two: (a) from the beginning of Islam introduction up to Islamic Kingdoms in Indonesia and (b) Islam in modern era, particularly during independent movement. Location-based helps user to find articles of events happening in certain areas. User can zoom it for more detailed information.

**Result and Discussion**

**Navigation**

Digitory is employed with two main navigations: location-based navigation and time-based navigation, located at its homepage, in order to help users browse articles either based on location or time, as shown in Figure 2 below.

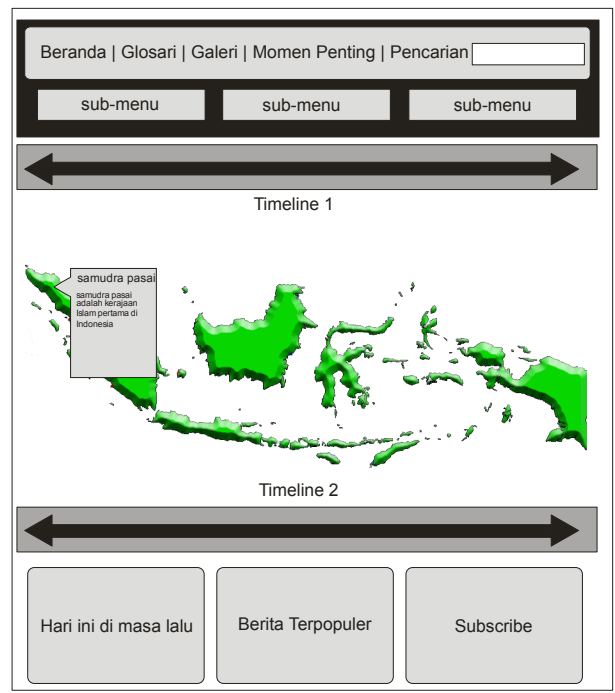

Figure **2.** Design of Digitory Homepage

1. Location-based navigation

This type of navigation is very helpful in understanding event based on its location. Users have the opportunity what events happened in Sumatra or Kalimantan (or other places) over periods of time, and therefore they can have whole and more general ideas of what happened in particular area.

User can click on certain location to see events occurred in that particular location. In order to see more events, user can zoom in the map and there are more events rendered for him. This navigation is great help for visual-type users.

2. Time-based navigation

Mostly time sequence plays important role in history. In order to help users understand orders of Islam-related events in Indonesia, these time-based navigations are provided: one at the top, showing history of Islam since its introduction to Indonesia until the glory of Islamic Kingdom; and one at the bottom for Islam role during national-independent movement..

User can scroll to the right in order to see events in more recent period, and to the left for older events. The navigation is populated with dots representing events occurred over period of time, which shows event descriptions if user hover over it and redirects to certain article if user clicks on it.

**Features**

1. Today on the past

This 'today on the past' section in homepage is populated with list of events happening in the same date with the date user access the web. This additional feature is important in giving knowledge of very important events happening in the past, with purpose of inspiring users to do great things on that day.

2. Related articles

As the user reads one article, Digitory smartly suggests other articles semantically related to it. In one section, the relation is based on location, either similar or nearby location. Another section is based on time, in which Digitory shows events occuring in similar dates, similar dates in different years, and in other dates close to it. However, the default is articles of events whose location and time are close to those of current articles.

3. Article-supporting gallery

In order to help users understand history, particular those of visual-type persons, gallery is made, providing images of events happening during Islam-related events in Indonesia.

Divided into two golden ration sections, Digitory's gallery page shows main image on its right side and gallery thumbnails on its left side. Thumbnails enable users to click on certain images and main image will change accordingly.

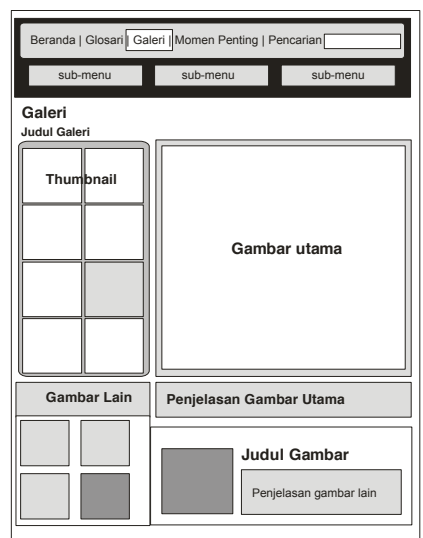

Figure 3. Gallery of images

**Conclusions**

Digitory is a supporting method for learning Islamic history in Indonesia, particularly for youth in modern era. Its time-based navigation allows user to search events in time sequence, in which articles of events are ordered from past to present. Additionally, it has location-based navigation, in form of map, giving access for users to browse articles of events based on their location of occuring.

For future development, it is suggested to complement Digitory content with more trusted sources of Islamic history in Indonesia.